\documentclass[11pt,a4paper]{article}
\pdfoutput=1

\usepackage{jheppub}
\usepackage[utf8]{inputenc}
\usepackage[T1]{fontenc}
\usepackage{lmodern}
\usepackage{microtype}
\usepackage{mathtools,bm}
\usepackage{mathrsfs}
\usepackage[nameinlink,capitalize]{cleveref}

\graphicspath{{./}}

\newcommand{\HP}{\mathrm{HP}}
\newcommand{\dd}{\mathrm{d}}
\newcommand{\F}{\mathcal{F}}
\newcommand{\E}{\mathcal{E}}
\newcommand{\Pmom}{\mathcal{P}}
\newcommand{\Mmom}{\mathcal{M}}
\newcommand{\Smin}{S_{\mathrm{D}}}
\newcommand{\Tmin}{T_{\mathrm{D}}}
\newcommand{\SHP}{S_{\HP}}
\newcommand{\THP}{T_{\HP}}
\newcommand{\wD}{w_{\mathrm{D}}}
\newcommand{\wHP}{w_{\HP}}

\newcommand{\sgn}{\operatorname{sgn}}

\newcommand{\Eq}[1]{Eq.(\ref{#1})}
\newcommand{\Eqs}[2]{Eqs.(\ref{#1})--(\ref{#2})}
\newcommand{\Fig}[1]{Fig.(\ref{#1})}
\newcommand{\SecRef}[1]{Sec.~\ref{#1}}

\title{\boldmath Hawking--Page Universality, Thermodynamic Dipoles and Categorical Defects}

\author[a]{Emilio Torrente-Lujan}
\affiliation[a]{F\'isica Te\'orica, Departamento de F\'isica, Universidad de Murcia, Campus de Espinardo, E-30100 Murcia, Spain}
\emailAdd{etl@um.es}

\abstract{We reconsider the Hawking--Page transition using the common thermodynamic vector field whose zeros include the Davies 
and Hawking--Page points.  In the elementary AdS branch their winding numbers are $w_{\rm D}=-1$ and $w_{\rm HP}=+1$, so the pair has zero total charge but a non-zero signed first moment.  After normalization by the Davies scales this moment gives the familiar universal ratios $C_S$ and $C_T$; in four dimensions $C_S=2$ and $C_T=2/\sqrt{3}-1$.  We check the construction for Schwarzschild--AdS, grand-canonical Reissner--Nordstr\"om--AdS, charged non-rotating black holes in arbitrary dimension, and Kerr--AdS at fixed angular velocity.  The same reduced geometry gives a barrier $B=1/3$ in four dimensions and $B(d)=1/[(d-1)(d-3)]$.  Finally we propose a formulation involving  a defect-resolved version for categorical or non-invertible symmetry sectors.}


\begin{document}
\maketitle

\section{Introduction}
\label{sec:introduction}

Black-hole thermodynamics bridges gravity and statistical mechanics.  The four laws of black-hole mechanics \citep{Bardeen1973}, the entropy formula of Bekenstein \citep{Bekenstein1973}, Hawking radiation \citep{Hawking1975}, and the Hawking--Page transition in anti-de Sitter space \citep{HawkingPage1983} are standard pieces of this bridge.  In holography the Hawking--Page transition is read as a confinement/deconfinement transition of the boundary gauge theory on a compact space \citep{Witten1998}.  Charged or rotating AdS black holes add the usual thermodynamic complications: ensemble dependence, metastable branches, swallow tails and extended phase diagrams \citep{Chamblin1999a,Chamblin1999b,Caldarelli1999,KubiznakMann2012,KubiznakMannTeo2017}.

There is, however, a rather simple numerical structure hidden in these phase diagrams.  Let $(\Smin,\Tmin)$ denote the point at which the temperature is minimal along the black-hole branch.  In the examples used below this is also the elementary Davies point, where the corresponding heat capacity diverges \citep{Davies1977}.  Let $(\SHP,\THP)$ denote the Hawking--Page point.  Then the ratios
\begin{eqnarray}
C_S &=& \frac{\SHP-\Smin}{\Smin}, \label{eq:CSCT-intro} \\
C_T &=& \frac{\THP-\Tmin}{\Tmin} \nonumber
\end{eqnarray}
were first isolated in Ref,\cite{Belhaj2020Universal}.  For four-dimensional non-rotating AdS black holes one obtains $C_S=2$ and $C_T=2/\sqrt{3}-1$.  Related universal structures, including a dual relation between minimum temperatures and Hawking--Page temperatures in neighbouring dimensions, were discussed in Ref.\cite{WeiLiuMann2020Dual}.

A second, more recent, line of work has described black-hole thermodynamics in topological terms.  In the off-shell free-energy landscape, black-hole branches can be treated as thermodynamic defects carrying integer winding numbers \citep{WeiLiuMann2022Topology}.  This idea has been used for Hawking--Page transitions, critical points, Born--Infeld and R\'enyi examples, and for broader topological classifications of black-hole families \citep{YerraBhamidipatiMukherji2022Topology,YerraBhamidipatiMukherji2023BornInfeld,BarziMoumniMasmar2024Renyi,FangJiangZhang2023,WeiLiuMann2024Universal,WuLiuWuMann2025Novel,ZhuLiuWu2025Rotating,ChenZhuWu2025BTZ}.  Especially useful for our purpose is the common vector field of \citet{HazarikaGogoiPhukon2025}.  Its zeros include both Davies-type points and Hawking--Page points.  In the elementary diagram the Davies point has $w_{\rm D}=-1$, whereas the Hawking--Page point has $w_{\rm HP}=+1$.

In this work we ask how the older universal ratios look in this topological language.  The answer is almost forced by the signs.  The two charges form a neutral pair,
\begin{eqnarray}
w_{\rm D}+w_{\rm HP} &=& 0,
\end{eqnarray}
but they do not sit at the same thermodynamic point.  Their first signed moments are
\begin{eqnarray}
\Pmom_S &=& \wHP\SHP+\wD\Smin \label{eq:dipole-intro} \\
&=& \SHP-\Smin, \nonumber \\
\Pmom_T &=& \wHP\THP+\wD\Tmin \nonumber \\
&=& \THP-\Tmin. \nonumber
\end{eqnarray}
Dividing by the Davies scales gives back \Eq{eq:CSCT-intro}.  Thus the constants $C_S$ and $C_T$ can be read as the entropy and temperature components of a thermodynamic topological dipole.

The winding numbers are genuine topological charges, whereas the components in \Eq{eq:dipole-intro} depend on the thermodynamic coordinates used to locate the zeros.  The point of the construction is therefore 
not to claim neccesarily a new homotopy invariant, but 
to keep in one notation the integer orientation of the defects and the scale-free thermodynamic separations.  In that sense the Hawking--Page constants become part of the same bookkeeping as the winding numbers.

The structure of the paper is the following.  In \SecRef{sec:thermo-defects} we review the common vector field and give the local sign rule for Hawking--Page and Davies zeros.  In \SecRef{sec:dipoles} we define the signed moments and show how $C_S$, $C_T$ and the normalized barrier depend only on the reduced shape of the free-energy curve.  We then compute the moments in four dimensions, in arbitrary spacetime dimension for charged non-rotating black holes, and perturbatively for Kerr--AdS at fixed angular velocity.  The barrier constant is
\begin{eqnarray}
B &=& \frac{\F(\Smin)}{\Tmin\Smin},
\end{eqnarray}
with $B=1/3$ in four dimensions and $B=1/[(d-1)(d-3)]$ in the higher-dimensional charged family.  Finally, we discuss two extensions: a multipole description for diagrams with more than one Davies/Hawking--Page pair, and a tentative categorical refinement based on defect-resolved partition functions.

Finally we propose another direction of study.
 Generalized global symmetries are implemented by topological operators \citep{GaiottoKapustinSeibergWillett2015}; non-invertible symmetries replace group-like multiplication by categorical fusion rules \citep{BhardwajBottiniSchaferNamekiTiwari2023,Shao2023TASI}; and several holographic or brane realizations are now known \citep{ApruzziBahBonettiSchaferNameki2023,GutperleLiRathoreRoumpedakis2024,AntinucciBeniniCopettiGalatiRizi2025}.  Since the Hawking--Page transition is sensitive to the topology of the Euclidean thermal circle, it is natural to conjecture whether inserting such defects shifts or splits the transition temperature.  In a first approximation, we  define quantities that can be calculated in models where the relevant defect sectors and both Euclidean saddles are under analytic control.

\section{Davies and Hawking--Page zeros }
\label{sec:thermo-defects}

Let us work in units in which $G=\hbar=c=k_B=1$.  In four-dimensional formulae it will often be convenient to use a reduced entropy $S=r_+^2$ and a reduced temperature differing from the physical Hawking temperature by an overall factor of $\pi$.  Such constant rescalings do not affect the ratios $C_S$, $C_T$, or the winding signs.

Let us consider a one-parameter black-hole branch at fixed ensemble variables $Y_i$, such as fixed pressure, fixed electric potential, or fixed angular velocity.  Let $\E(S;Y_i)$ be the appropriate Legendre-transformed energy whose derivative gives the ensemble temperature,
\begin{eqnarray}
T(S;Y_i) &=& \left(\frac{\partial \E}{\partial S}\right)_{Y_i}. \label{eq:T-from-E}
\end{eqnarray}
For example, $\E=M$ in a canonical neutral ensemble, while $\E=M-\Phi Q$ or $\E=M-\Omega J$ in the corresponding grand-canonical ensembles.  The on-shell free energy is
\begin{eqnarray}
\F(S;Y_i) &=& \E(S;Y_i)-T(S;Y_i)S. \label{eq:on-shell-free-energy}
\end{eqnarray}
Along the black-hole branch,
\begin{eqnarray}
\left(\frac{\partial \F}{\partial S}\right)_{Y_i} &=& -S\left(\frac{\partial T}{\partial S}\right)_{Y_i}. \label{eq:F-derivative}
\end{eqnarray}
The Davies-type point considered in this work is the minimum-temperature point,
\begin{eqnarray}
\left(\frac{\partial T}{\partial S}\right)_{Y_i} &=& 0, \label{eq:Davies-definition} \\
\left(\frac{\partial^2T}{\partial S^2}\right)_{Y_i} &>& 0, \nonumber
\end{eqnarray}
where the heat capacity $C_Y=T(\partial S/\partial T)_Y$ diverges.  We denote it by $(\Smin,\Tmin)$.  The Hawking--Page point is defined by
\begin{eqnarray}
\F(\SHP;Y_i) &=& 0, \label{eq:HP-definition} \\
\THP &=& T(\SHP;Y_i), \nonumber
\end{eqnarray}
where the black-hole saddle and thermal AdS have equal free energy.

\begin{figure}[tbph]
\centering
\includegraphics[width=0.92\textwidth]{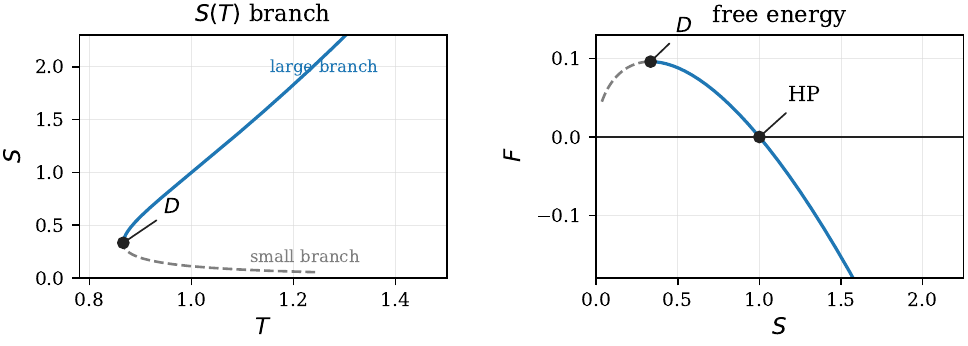}
\caption{Reduced Schwarzschild--AdS phase schematic portrait (adapted from \cite{Belhaj2020Universal}).  Left: the Davies point $D$ is the minimum of $T(S)$ and separates the small and large branches.  Right: the same point is the maximum of $\F(S)$, while HP is the zero of $\F$ on the stable large-black-hole branch.  Dashed portions denote locally unstable states.}
\label{fig:legacy}
\end{figure}

The basic picture is shown in \Fig{fig:legacy}.  
The same point that minimizes $T(S)$ is the maximum of the on-shell free energy, and the Hawking--Page transition occurs where that free energy crosses zero.

Following the common-field construction of \citep{HazarikaGogoiPhukon2025}, define a two-component vector field on the auxiliary space $(S,\theta)$ by
\begin{eqnarray}
\bm{\varphi}(S,\theta) &=& \left(\varphi^S,\varphi^\theta\right) \label{eq:common-vector-field} \\
&=& \left(\frac{1}{S}\frac{\partial \F^2}{\partial S},-\cot\theta\,\csc\theta\right). \nonumber
\end{eqnarray}
Equivalently, if the thermodynamic functions are more naturally expressed in terms of the horizon radius $r_+$, one may use
\begin{eqnarray}
\bm{\varphi}(r_+,\theta) &=& \left(\frac{1}{r_+}\frac{\partial \F^2}{\partial r_+},-\cot\theta\,\csc\theta\right). \label{eq:common-vector-field-r}
\end{eqnarray}
The second component vanishes only at $\theta=\pi/2$.  The first component satisfies
\begin{eqnarray}
\varphi^S &=& \frac{2\F}{S}\frac{\partial \F}{\partial S} \label{eq:phiS-factorization} 
= -2\F\frac{\partial T}{\partial S} 
= -2\F\frac{T}{C_Y}, 
\end{eqnarray}
where \Eq{eq:F-derivative} was used.  Thus the isolated zeros of \Eq{eq:common-vector-field} are precisely the points at which
\begin{eqnarray}
\F &=& 0 \quad \text{or}\quad \frac{1}{C_Y}=0, \label{eq:zeros-common-field}
\end{eqnarray}
namely Hawking--Page or Davies-type points.

A consequence of \Eq{eq:F-derivative} is worth isolating, since it underlies the barrier of \SecRef{subsec:barrier} and the sign rule below.  Because $\F_S=-S\,T_S$, the stationary points of the on-shell free energy coincide with those of the temperature: every Davies point is an extremum of $\F$.  Differentiating once more gives $\F_{SS}=-(T_S+S\,T_{SS})$, so at a Davies point ($T_S=0$) one has $\F_{SS}=-S\,T_{SS}$.  A \emph{minimum} of the temperature is therefore a \emph{maximum} of the free energy.  In the standard portrait of \Fig{fig:legacy} the Davies point is exactly the top of the free-energy barrier separating thermal AdS from the black hole.

The winding number of a zero $z_i$ is
\begin{eqnarray}
w_i &=& \frac{1}{2\pi}\oint_{\Gamma_i}\dd\Omega, \label{eq:winding-number} 
\end{eqnarray}
where
\begin{eqnarray}
\Omega &=& \arctan\frac{n^2}{n^1}, \nonumber \\
n^a &=& \frac{\varphi^a}{\sqrt{\varphi^b\varphi^b}}, \nonumber
\end{eqnarray}
and where $\Gamma_i$ is a small positively oriented contour around the zero.  For a simple zero this reduces to the sign of the Jacobian,
\begin{eqnarray}
w_i &=& \sgn\left[\det\left(\frac{\partial \varphi^a}{\partial x^b}\right)_{z_i}\right], \label{eq:winding-jacobian} \\
x^b &=& (S,\theta). \nonumber
\end{eqnarray}

For the elementary branch the local charge assignment is simple.  Assume a single locally unstable small-black-hole branch followed by a stable large-black-hole branch.  If the Davies point is a local minimum of $T(S)$ with $\F(\Smin)>0$, and the Hawking--Page point lies on the stable branch with $(\partial T/\partial S)_{\SHP}>0$, then
\begin{eqnarray}
\wD &=& -1, \label{eq:charges-result} \\
\wHP &=& +1. \nonumber
\end{eqnarray}
The check is completely local.  Near $\theta=\pi/2+\delta\theta$, one has 
$$-\cot\theta\csc\theta=\delta\theta+O(\delta\theta^3),$$
 so the angular Jacobian contributes a positive sign.  
 At the Davies point, $T_S=0$ and $T_{SS}>0$.  Differentiating \Eq{eq:phiS-factorization} gives
\begin{eqnarray}
\left.\frac{\partial \varphi^S}{\partial S}\right|_{\mathrm{D}} &=& -2\F(\Smin)T_{SS}(\Smin)<0. \label{eq:jac-davies}
\end{eqnarray}
Thus $w_{\mathrm{D}}=-1$.  At the Hawking--Page point, $\F=0$ and $T_S>0$.  Using $\F_S=-S T_S$,
\begin{eqnarray}
\left.\frac{\partial \varphi^S}{\partial S}\right|_{\HP} &=& -2\F_S(\SHP)T_S(\SHP) \label{eq:jac-hp} \\
&=& 2\SHP\left[T_S(\SHP)\right]^2>0, \nonumber
\end{eqnarray}
and therefore $w_{\HP}=+1$.  The same computation gives the winding number of any simple transition zero in terms of purely local data, which is what one needs for phase diagrams with several Davies and Hawking--Page points.

It is useful to record the same calculation as a local sign rule.  At a simple zero of the common field \Eq{eq:common-vector-field},
we have
\begin{eqnarray}
\wHP&=&+1, 
\end{eqnarray}
for a simple Hawking--Page zero with $T_S\neq0$,
and
\begin{eqnarray}
\wD&=&-\,\sgn\F(\Smin)\,\sgn T_{SS}(\Smin),
\end{eqnarray}
for a simple Davies zero.

The angular factor is the same positive factor as above.  At a Hawking--Page zero $\F=0$, so 
$$\partial_S\varphi^S=-2\F_S T_S=2S\,T_S^2>0$$ 
whenever $T_S\neq0$.  At a Davies zero $T_S=0$, so $\partial_S\varphi^S=-2\F\,T_{SS}$, whose sign 
is $$-\sgn\F\,\sgn T_{SS}.$$

Thus a Davies point carries the usual charge $-1$ only when it is a temperature minimum on the high-free-energy side ($\F>0$), and it flips to $+1$ when it sits in the black-hole-dominated region ($\F<0$) or becomes a temperature maximum, as happens along reentrant or multi-branch curves.  The charge configuration of an arbitrary phase diagram can therefore be read directly from the signs of $\F$ and $T_{SS}$ at its transition points, with no Jacobian evaluation.  A non-simple zero, at which a Davies and a Hawking--Page point merge ($\F=0$ and $T_S=0$ simultaneously), is precisely where two charges annihilate and the dipole degenerates.

\begin{figure}[tbp]
\centering
\includegraphics[width=0.66\textwidth]{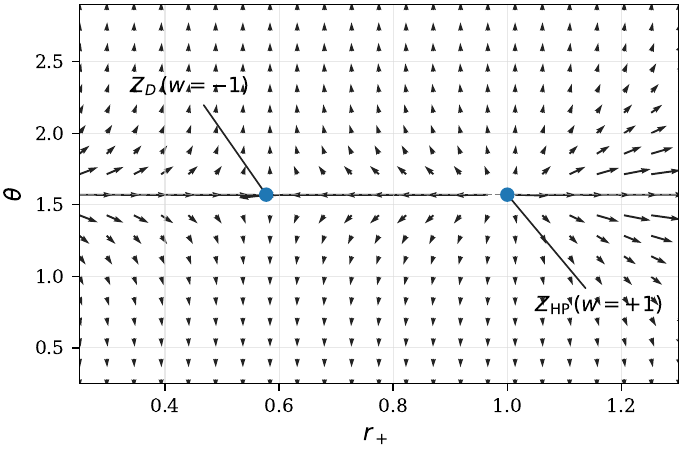}
\caption{Normalized common vector field for Schwarzschild--AdS in the $(r_+,\theta)$ plane.  The zero at $r_+=1/\sqrt{3}$ is the Davies point ($w=-1$), while the zero at $r_+=1$ is the Hawking--Page point ($w=+1$).}
\label{fig:vector-field}
\end{figure}

For Schwarzschild--AdS the horizon-radius version of the field gives
\begin{eqnarray}
\varphi^{r_+} &=& \frac{1}{8}\left(3r_+^4-4r_+^2+1\right), \label{eq:sads-vector} \\
\varphi^\theta &=& -\cot\theta\,\csc\theta, \nonumber
\end{eqnarray}
with zeros at $(r_+,\theta)=(1/\sqrt{3},\pi/2)$ and $(1,\pi/2)$.  The normalized vector portrait is shown in \Fig{fig:vector-field}.  The first zero is the Davies point, with charge $-1$, and the second is the Hawking--Page point, with charge $+1$.

\section{Universal ratios as signed moments}
\label{sec:dipoles}

Let us now connect the topological charges to the universal ratios.  Let $z_i$ denote isolated Hawking--Page or Davies zeros of the common field, and let $w_i$ be their winding numbers.  For any thermodynamic coordinate $X$ evaluated at the zeros, define the signed first moment
\begin{eqnarray}
\Pmom_X &=& \sum_i w_i X_i. \label{eq:first-moment-general}
\end{eqnarray}

For a single Davies/Hawking--Page pair,
\begin{eqnarray}
Q_{\mathrm{tot}} &\equiv& w_{\mathrm{D}}+w_{\HP} \label{eq:neutral-pair} \\
&=& 0, \nonumber
\end{eqnarray}
so the pair is neutral. 
 Nevertheless its first moments are non-zero:
\begin{eqnarray}
\Pmom_S &=& \SHP-\Smin, \label{eq:single-pair-moments} \\
\Pmom_T &=& \THP-\Tmin. \nonumber
\end{eqnarray}

For later reference we define the normalized dipole ratios of a single Hawking--Page/Davies pair, with charges $\wHP=+1$ and $\wD=-1$, by
\begin{eqnarray}
\mathfrak{C}_S &=& \frac{\Pmom_S}{\Smin}, \label{eq:topological-dipole-ratios} \\
\mathfrak{C}_T &=& \frac{\Pmom_T}{\Tmin}. \nonumber
\end{eqnarray}

With this notation,
\begin{eqnarray}
\mathfrak{C}_S &=& C_S, \qquad \label{eq:dipole-equals-universal} 
\mathfrak{C}_T = C_T. 
\end{eqnarray}
In this precise, coordinate-level sense, the  constants $C_S$ and $C_T$ are the two components of the normalized Davies/Hawking--Page dipole.  The individual ratios and the winding numbers are known from earlier work.  They are two components of a single moment hierarchy.

Note that the distinction between topological charge and thermodynamic moment is essential.  The integers $w_i$ are invariant under smooth deformations that do not create, annihilate, or merge zeros, whereas the coordinates $S_i$ and $T_i$ are ordinary thermodynamic data.  The ratios \Eq{eq:topological-dipole-ratios} are therefore mixed objects: topological in orientation and thermodynamic in magnitude.  Their universality in the examples below comes from the cancellation of ensemble scales, not from homotopy alone.

\begin{figure}[tbp]
\centering
\includegraphics[width=0.92\textwidth]{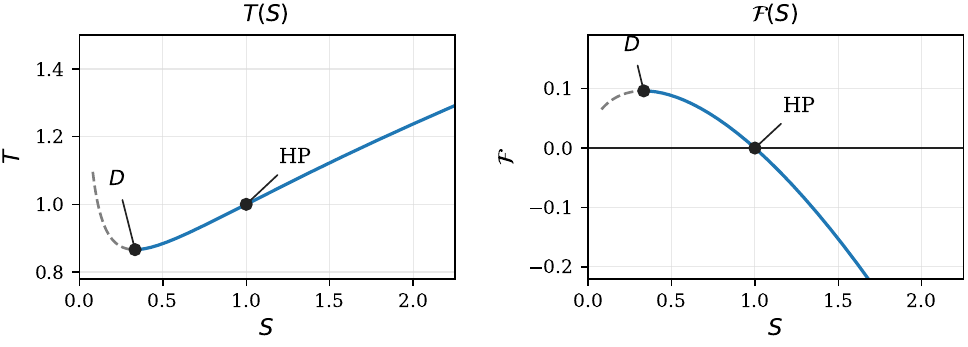}
\caption{Four-dimensional Schwarzschild--AdS branch plotted as $T(S)$ and $\F(S)$.  The oriented separation from $D$ to HP gives the two dipole components, $\Pmom_S=\SHP-\Smin$ and $\Pmom_T=\THP-\Tmin$.}
\label{fig:sads-curves}
\end{figure}

The \Fig{fig:sads-curves} shows the same statement on the Schwarzschild--AdS curves.  The sign of the winding number fixes the orientation of the separation from $D$ to HP.

It is useful to introduce also the ratio variables
\begin{eqnarray}
R_S &\equiv& \frac{\SHP}{\Smin} \label{eq:R-ratios} 
= 1+C_S, \nonumber \\
R_T &\equiv& \frac{\THP}{\Tmin} \nonumber 
= 1+C_T. \nonumber
\end{eqnarray}
The four-dimensional non-rotating result $C_S=2$ and $C_T=2/\sqrt{3}-1$ is equivalently
\begin{eqnarray}
R_S &=& 3, \label{eq:R-universal-fourD} \\
R_T &=& \frac{2}{\sqrt{3}}. \nonumber
\end{eqnarray}

Let us explore structural reason for the universality seen in the examples below.  Suppose the reduced temperature of a black-hole family has the scaling form
\begin{eqnarray}
T(S) &=& \tau\,t\!\left(\frac{S}{S_0}\right), \label{eq:shape-form}
\end{eqnarray}
for some amplitude $\tau$, scale $S_0$, and dimensionless shape function $t(x)$.  Then $T_S\propto t'(S/S_0)$, so the Davies point sits at $\Smin=x_{\mathrm D}S_0$ with $x_{\mathrm D}$ a root of $t'$.  Integrating $\F_S=-S\,T_S$ gives $\F(S)=\tau S_0\,f(S/S_0)$ with
\begin{eqnarray}
f(x) &=& \int_0^x t(x')\,\dd x'-x\,t(x), \label{eq:shape-f}
\end{eqnarray}
so the Hawking--Page point sits at $\SHP=x_{\HP}S_0$ with $x_{\HP}$ a root of $f$.  

Both $x_{\mathrm D}$ and $x_{\HP}$ are pure numbers fixed by the shape $t$ alone, and hence
\begin{eqnarray}
C_S &=& \frac{x_{\HP}}{x_{\mathrm D}}-1, \label{eq:shape-invariants} \\
C_T &=& \frac{t(x_{\HP})}{t(x_{\mathrm D})}-1, \nonumber \\
B &=& \frac{f(x_{\mathrm D})}{x_{\mathrm D}\,t(x_{\mathrm D})}, \nonumber
\end{eqnarray}
are independent of $\tau$ and $S_0$, and the last of these is the nucleation-barrier ratio used in \SecRef{subsec:barrier}.

This gives the working criterion used below: if two black-hole branches have reduced temperatures of the form \Eq{eq:shape-form} with the same shape function $t$, then they have identical $C_S$, $C_T$, and $B$, regardless of their amplitudes and scales.

The four-dimensional non-rotating family realizes a single shape, $t(x)=(1+3x)/(4\sqrt{x})$: for Reissner--Nordstr\"om--AdS the choice $S_0=\sigma$ in \Eq{eq:rn-thermo} returns exactly this $t$, which is why the constants do not depend on the electric potential.  In this language the ratios are invariants of a ``corresponding-states'' shape, and universality is broken precisely when a deformation changes the shape itself.  Rotation does precisely this, and produces the corrections computed in \SecRef{sec:kerr}.

\section{Four-dimensional non-rotating AdS black holes}
\label{sec:fourD}

\subsection{Schwarzschild--AdS}
\label{subsec:sads}

For four-dimensional Schwarzschild--AdS, set the AdS radius to unity and use the reduced entropy $S=r_+^2$.  The mass, temperature, and free energy are
\begin{eqnarray}
M &=& \frac{\sqrt{S}(1+S)}{2}, \label{eq:sads-thermo} \\
T &=& \frac{1+3S}{4\sqrt{S}}, \nonumber \\
\F &=& \frac{\sqrt{S}}{4}(1-S). \nonumber
\end{eqnarray}
The Davies point is obtained from $T_S=0$,
\begin{eqnarray}
\Smin &=& \frac{1}{3}, \label{eq:sads-D} \\
\Tmin &=& \frac{\sqrt{3}}{2}. \nonumber
\end{eqnarray}
The Hawking--Page point is obtained from $\F=0$,
\begin{eqnarray}
\SHP &=& 1, \label{eq:sads-HP} \\
\THP &=& 1. \nonumber
\end{eqnarray}
Therefore
\begin{eqnarray}
\Pmom_S &=& \SHP-\Smin \label{eq:sads-moments} \\
= \frac{2}{3},  \\
\Pmom_T &=& \THP-\Tmin 
= 1-\frac{\sqrt{3}}{2}, 
\end{eqnarray}
with normalized components
\begin{eqnarray}
C_S &=& \frac{\Pmom_S}{\Smin} \label{eq:sads-CSCT} 
= 2, \\
C_T &=& \frac{\Pmom_T}{\Tmin} 
= \frac{2}{\sqrt{3}}-1. \nonumber
\end{eqnarray}

In extended thermodynamics, where $P=3/(8\pi L^2)$, one instead obtains
\begin{eqnarray}
\Smin &=& \frac{1}{8\pi P}, \label{eq:sads-extended} \\
\SHP &=& \frac{3}{8\pi P}, \nonumber 
\end{eqnarray}
and
\begin{eqnarray}
\Tmin &=& \sqrt{2\pi P}, \nonumber \\
\THP &=& \sqrt{\frac{8\pi P}{3}}. \nonumber
\end{eqnarray}
The pressure cancels from $C_S$ and $C_T$, so the same dipole components \Eq{eq:sads-CSCT} are recovered.

\subsection{Reissner--Nordstr\"om--AdS at fixed electric potential}
\label{subsec:rnadS}

For four-dimensional Reissner--Nordstr\"om--AdS in the grand-canonical ensemble, let
\begin{eqnarray}
\sigma &\equiv& 1-\Phi^2, \label{eq:sigma-def} ,\qquad
0 < \sigma\leq 1, 
\end{eqnarray}
where $\Phi$ is the electric potential at infinity.  The reduced temperature and grand potential are
\begin{eqnarray}
T &=& \frac{3S+\sigma}{4\sqrt{S}}, \label{eq:rn-thermo} \\
\F &=& \frac{\sqrt{S}}{4}(\sigma-S). \nonumber
\end{eqnarray}
The Davies and Hawking--Page points are
\begin{eqnarray}
(\Smin,\Tmin) &=& \left(\frac{\sigma}{3},\frac{\sqrt{3\sigma}}{2}\right), \label{eq:rn-D} \\
(\SHP,\THP) &=& \left(\sigma,\sqrt{\sigma}\right). \label{eq:rn-HP}
\end{eqnarray}
The corresponding signed moments are
\begin{eqnarray}
\Pmom_S &=& \frac{2\sigma}{3}, \label{eq:rn-moments} \\
\Pmom_T &=& \sqrt{\sigma}\left(1-\frac{\sqrt{3}}{2}\right). \nonumber
\end{eqnarray}
After normalization by the negative-charge point, the scale $\sigma$ cancels:
\begin{eqnarray}
C_S &=& 2, \label{eq:rn-CSCT} \\
C_T &=& \frac{2}{\sqrt{3}}-1. \nonumber
\end{eqnarray}
As a consequence, the same charged dipole gives a parameter-independent result throughout the grand-canonical non-rotating family: the electric potential enters only through the common scale $\sigma$, which then cancels in both normalized components.

\subsection{The Hawking--Page nucleation barrier}
\label{subsec:barrier}

By the local sign rule the Davies point is the maximum of the on-shell free energy, i.e.\ the top of the barrier separating the thermal-AdS phase ($\F\to0$ as $S\to0$) from the stable black hole.  Its height $\F(\Smin)$ is the quantity that controls the rate of the Hawking--Page transition in the free-energy-landscape treatment of the kinetics \citep{WeiLiuMann2022Topology,LiWang2020Kinetics}.  It is natural to measure this height against the energy scale set by the same defect, $\Tmin\Smin$, and to define the dimensionless barrier
\begin{eqnarray}
B &=& \frac{\F(\Smin)}{\Tmin\,\Smin}. \label{eq:barrier-def}
\end{eqnarray}
For four-dimensional Schwarzschild--AdS, \Eq{eq:sads-thermo}--\Eq{eq:sads-D} give $\F(\Smin)=1/(6\sqrt{3})$ and $\Tmin\Smin=1/(2\sqrt{3})$, so
\begin{eqnarray}
B &=& \frac{1}{3}. \label{eq:barrier-4d}
\end{eqnarray}
The grand-canonical Reissner--Nordstr\"om--AdS family gives the same value: from \Eq{eq:rn-thermo}--\Eq{eq:rn-HP}, $\F(\Smin)=\sigma^{3/2}/(6\sqrt{3})$ and $\Tmin\Smin=\sqrt{3}\,\sigma^{3/2}/6$, and the scale $\sigma$ cancels to leave $B=1/3$.  By the shape argument this is no accident: $B$ is a functional of the four-dimensional shape function and is therefore shared by the entire non-rotating family.  The barrier provides a third universal constant, of energy type, alongside the entropy and temperature dipole components $C_S$ and $C_T$, and it connects the topology directly to the transition kinetics.

\section{Higher-dimensional charged AdS black holes}
\label{sec:higherD}

The same construction carries over, with little more than a change of exponents, to the charged non-rotating AdS family in $d$ spacetime dimensions, again worked out in the grand-canonical ensemble.  Let $\omega_{d-2}$ be the volume of the unit $(d-2)$-sphere and let $\phi$ be the electric potential.  Define
\begin{eqnarray}
A_d(\phi) &\equiv& (d-2)-2(d-3)\phi^2, \label{eq:Adphi} \\
A_d(\phi) &>& 0. \nonumber
\end{eqnarray}
The Davies and Hawking--Page entropies are
\begin{eqnarray}
\Smin &=& \frac{\omega_{d-2}}{4} \left[\frac{d-3}{(d-2)(d-1)}\right]^{(d-2)/2} \left[A_d(\phi)\right]^{(d-2)/2}, \label{eq:higher-SD} \\
\SHP &=& \frac{\omega_{d-2}}{4(d-2)^{(d-2)/2}} \left[A_d(\phi)\right]^{(d-2)/2}. \label{eq:higher-SH}
\end{eqnarray}
The corresponding temperatures are
\begin{eqnarray}
\Tmin &=& \frac{\sqrt{(d-3)(d-1)}}{2\pi\sqrt{d-2}} \sqrt{A_d(\phi)}, \label{eq:higher-TD} \\
\THP &=& \frac{\sqrt{d-2}}{2\pi} \sqrt{A_d(\phi)}. \label{eq:higher-TH}
\end{eqnarray}

Hence the winding-normalized dipole components are
\begin{eqnarray}
C_S(d) &=& \frac{\SHP-\Smin}{\Smin} \label{eq:higher-CS} \\
&=& \left(\frac{d-1}{d-3}\right)^{(d-2)/2}-1, \nonumber \\
C_T(d) &=& \frac{\THP-\Tmin}{\Tmin} \label{eq:higher-CT} \\
&=& \frac{d-2}{\sqrt{(d-1)(d-3)}}-1. \nonumber
\end{eqnarray}
All dependence on the electric potential cancels.  The topological charge assignment remains $\wD=-1$, $\wHP=+1$, while the thermodynamic distance between the charges depends on the spacetime dimension.
The full dependence is carried by \Eqs{eq:higher-CS}{eq:higher-CT}, and the trend is read off directly: the temperature component falls quickly with $d$, whereas the entropy component approaches a finite limit.  Indeed,
\begin{eqnarray}
\lim_{d\to\infty}C_T(d) &=& 0, \label{eq:large-d-limits} \\
\lim_{d\to\infty}C_S(d) &=& e-1. \nonumber
\end{eqnarray}

In the dipole language, higher dimensions squeeze the temperature separation of the two charged defects while leaving a finite entropy separation.

The same construction fixes the nucleation barrier of \SecRef{subsec:barrier} in every dimension.  Reconstructing $\F(S)$ from the $d$-dimensional reduced temperature and evaluating \Eq{eq:barrier-def} at the Davies point gives
\begin{eqnarray}
B(d) &=& \frac{1}{(d-1)(d-3)}, \label{eq:barrier-d}
\end{eqnarray}
independently of the electric potential.  Equivalently 
$$B(d)=(1+C_T(d))^2/(d-2)^2,$$
 so the barrier and the temperature dipole component are tied together throughout the charged non-rotating family.  As $d\to\infty$ the barrier closes as $B(d)\sim 1/d^2$, consistent with the vanishing of the temperature separation $C_T(d)\to0$ while the entropy separation stays finite.  The result is plotted in \Fig{fig:barrier}.

\setcounter{figure}{4}
\begin{figure}[tbp]
\centering
\includegraphics[width=0.94\textwidth]{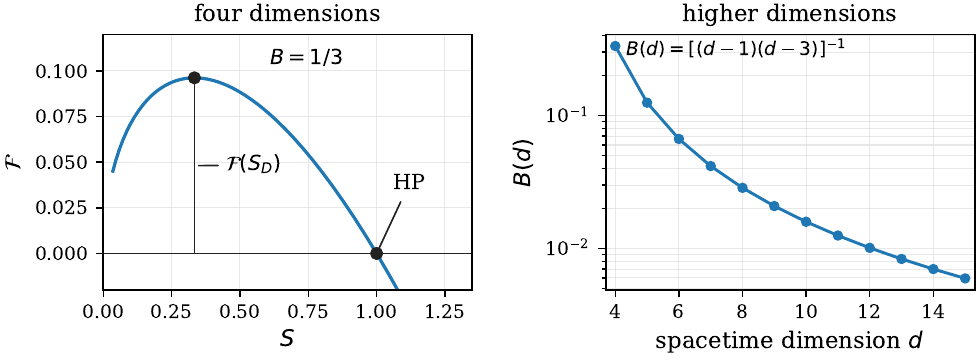}
\caption{Normalized Hawking--Page barrier.  In four dimensions the Davies point is the top of the free-energy barrier, giving $B=\F(\Smin)/(\Tmin\Smin)=1/3$.  In the charged non-rotating family the higher-dimensional result is $B(d)=1/[(d-1)(d-3)]$.}
\label{fig:barrier}
\end{figure}

\section{Kerr--AdS at fixed angular velocity}
\label{sec:kerr}

Rotation introduces an additional scale, and one should not expect exact parameter independence to survive.  Following the grand-canonical thermodynamic treatment of Kerr--AdS black holes \citep{Caldarelli1999}, in four dimensions at fixed angular velocity $\Omega$ and AdS radius $L=1$, the mass, temperature, and grand potential may be written in terms of the reduced entropy $S$ and $\Omega$ as
\begin{eqnarray}
M^2 &=& \frac{S(1+S)^3}{4(1+S-S\Omega^2)}, \label{eq:kerr-M}
\end{eqnarray}
\begin{eqnarray}
T(S,\Omega) &=& \sqrt{\frac{S(1+S)}{1+S-S\Omega^2}} \frac{1-2S(\Omega^2-2)-3S^2(\Omega^2-1)}{4S(1+S)}, \label{eq:kerr-T}
\end{eqnarray}
\begin{eqnarray}
\F(S,\Omega) &=& \frac{\sqrt{\dfrac{S(1+S)}{1+S-S\Omega^2}}\left[1-S^2(1-\Omega^2)\right]}{4(1+S)}. \label{eq:kerr-F}
\end{eqnarray}

The Hawking--Page entropy and temperature are
\begin{eqnarray}
\SHP &=& \frac{1}{\sqrt{1-\Omega^2}}, \label{eq:kerr-HP} \\
\THP &=& \frac{1+\sqrt{1-\Omega^2}}{2}, \nonumber \\
|\Omega| &<& 1. \nonumber
\end{eqnarray}
The Davies entropy is the positive root that continuously approaches $1/3$ as $\Omega\to0$ of
\begin{eqnarray}
3(1-\Omega^2)^2\Smin^4 +4(1-\Omega^2)(2-\Omega^2)\Smin^3 +6(1-\Omega^2)\Smin^2 -1 &=& 0. \label{eq:kerr-D-polynomial}
\end{eqnarray}
For small angular velocity,
\begin{eqnarray}
\Smin &=& \frac{1}{3}+\frac{\Omega^2}{6}+\frac{23\Omega^4}{192}+O(\Omega^6), \label{eq:kerr-SD-expansion} \\
\Tmin &=& \frac{\sqrt{3}}{2}\left(1-\frac{\Omega^2}{4}-\frac{7\Omega^4}{128}+O(\Omega^6)\right), \label{eq:kerr-TD-expansion} \\
\SHP &=& 1+\frac{\Omega^2}{2}+\frac{3\Omega^4}{8}+O(\Omega^6), \label{eq:kerr-SH-expansion} \\
\THP &=& 1-\frac{\Omega^2}{4}-\frac{\Omega^4}{16}+O(\Omega^6). \label{eq:kerr-TH-expansion}
\end{eqnarray}
Therefore the dipole components become
\begin{eqnarray}
C_S(\Omega) &=& 2+\frac{3\Omega^4}{64}+O(\Omega^6), \label{eq:kerr-CS} \\
C_T(\Omega) &=& \frac{2}{\sqrt{3}}-1-\frac{\sqrt{3}}{192}\Omega^4+O(\Omega^6). \label{eq:kerr-CT}
\end{eqnarray}
The $O(\Omega^2)$ terms cancel in the normalized ratios, so the leading rotational deformation appears only at fourth order.  The winding numbers themselves remain unchanged, $\wD=-1$ and $\wHP=+1$, throughout the regular fixed-$\Omega$ regime with $|\Omega|<1$.  Rotation therefore deforms the thermodynamic length of the dipole without changing its topological orientation.

The nucleation barrier behaves in the same way.  Evaluating \Eq{eq:barrier-def} on the rotating solution \Eq{eq:kerr-T}--\Eq{eq:kerr-F} at the Davies entropy \Eq{eq:kerr-SD-expansion} gives
\begin{eqnarray}
B(\Omega) &=& \frac{1}{3}+O(\Omega^4), \label{eq:barrier-kerr}
\end{eqnarray}
so the barrier, like both normalized dipole components, has no $O(\Omega^2)$ correction.  At this order the rotation changes the individual Davies and Hawking--Page data, but the normalized separations are unchanged.

\section{Multipoles for richer phase diagrams}
\label{sec:multipoles}

The previous sections dealt a simple case: one Davies zero and one Hawking--Page zero.  Many black-hole phase diagrams are not that simple.  Reentrant transitions, higher-curvature terms, cavities, Born--Infeld electrodynamics, R\'enyi entropy, and Lovelock or Gauss--Bonnet examples may contain several heat-capacity divergences and several free-energy crossings \citep{Altamirano2013Reentrant,Frassino2014Multiple,SuLi2022Cavity,YerraBhamidipatiMukherji2023BornInfeld,HuCuiXu2024Reentrant,BarziMoumniMasmar2024Renyi}.  In such cases a single pair of numbers like $C_S$ and $C_T$ cannot be the whole story.

Let
\begin{eqnarray}
\mathscr{Z} &=& \{z_i \label{eq:defect-set} 
= (X_i,w_i)\,|\,\bm\varphi(z_i) 
= 0\} \nonumber
\end{eqnarray}
be the set of transition zeros in some chosen thermodynamic coordinate $X$.  The coordinate might be entropy, temperature, pressure, electric potential, or another variable adapted to the ensemble.  The total charge is
\begin{eqnarray}
Q_0 &\equiv& \sum_{z_i\in\mathscr{Z}} w_i. \label{eq:total-transition-charge}
\end{eqnarray}
When $Q_0=0$, the first signed moment
\begin{eqnarray}
\Pmom_X &=& \sum_{z_i\in\mathscr{Z}}w_iX_i \label{eq:multipole-first}
\end{eqnarray}
is independent of a shift of the origin $X\to X+X_0$.  A convenient scale is the centroid of the negative charges,
\begin{eqnarray}
X_- &\equiv& \frac{\sum_{w_i<0}|w_i|X_i}{\sum_{w_i<0}|w_i|}, \label{eq:negative-centroid}
\end{eqnarray}
which gives the normalized first moment
\begin{eqnarray}
C^{(1)}_X &=& \frac{\Pmom_X}{X_-}. \label{eq:first-multipole-normalized}
\end{eqnarray}
For a single Davies/Hawking--Page pair, this reduces to the ratios already discussed.  For several pairs, it measures the net oriented displacement between negative- and positive-winding transition points.

One may also record
\begin{eqnarray}
\Mmom_X^{(n)} &=& \sum_{z_i\in\mathscr{Z}}w_i\left(\frac{X_i}{X_-}\right)^n, \label{eq:higher-multipoles} \\
n &=& 1,2,\ldots . \nonumber
\end{eqnarray}
The terminology is borrowed from electrostatics, but only as an analogy: the ``charges'' are winding numbers and their positions are thermodynamic coordinates.  The use of this language is practical.  A total winding number can be too coarse.  Even a first moment can miss real differences between diagrams.

\begin{figure}[tbp]
\centering
\includegraphics[width=0.94\textwidth]{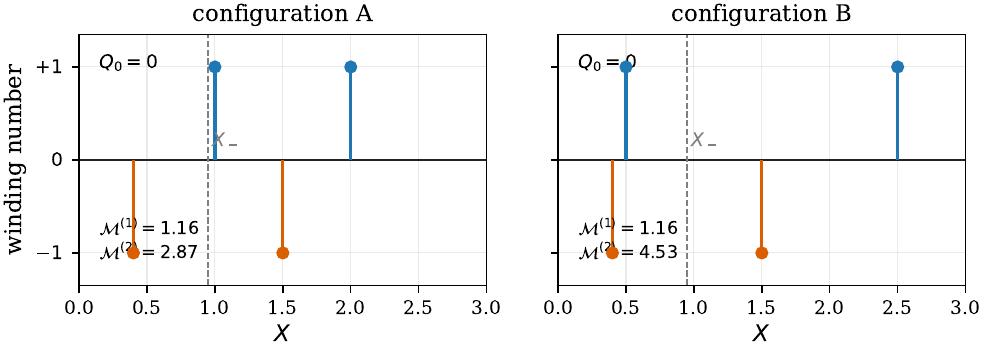}
\caption{Toy defect configurations illustrating the need for higher moments.  The two panels have the same total winding number and the same normalized first moment, but their second moments differ.  The dashed line marks the negative-charge centroid $X_-$.}
\label{fig:multipole}
\end{figure}

we illustrate in \Fig{fig:multipole} ithe point with a toy configuration of two positive and two negative defects.  Both panels have $Q_0=0$ and the same $\Mmom^{(1)}$, but they are separated by $\Mmom^{(2)}$.  This is the sense in which a moment spectrum refines the usual topological classification.  The total charge fixes the monopole data, and the Hawking--Page constants fix the simplest dipole data.  Higher moments then keep track of how the transition points are ordered along the branch.

For an explicit black-hole solution the procedure is straightforward.  One constructs the common field \Eq{eq:common-vector-field} or \Eq{eq:common-vector-field-r}, locates the Hawking--Page and Davies zeros in the ensemble under consideration, assigns their winding numbers, and then computes \Eq{eq:first-multipole-normalized} and \Eq{eq:higher-multipoles}.  The sign rule \Eq{eq:sign-rule} is useful here because it allows the charge of each simple zero to be read directly from the local shape of $T(S)$ and $\F(S)$.

\section{A categorical approach}
\label{sec:categorical}

The topology used above is a topology of thermodynamic state space, whose defects are zeros of a vector field in an auxiliary $(S,\theta)$ plane.  Categorical and non-invertible symmetries are different objects: their defects live in spacetime, or in the associated symmetry topological field theory.  Still, there is a natural place where the two languages may meet.  The Hawking--Page transition compares two Euclidean saddles with different thermal-circle topology, and line or surface defects can be sensitive to that difference.

In the boundary gauge theory the Hawking--Page transition is the large-$N$ deconfinement transition \citep{Witten1998}.  The Polyakov loop is a standard diagnostic.  In thermal AdS the Euclidean thermal circle is non-contractible, while in the Euclidean black-hole saddle it caps off at the horizon.  Thus a defect wrapping or linking the thermal circle can distinguish the two saddles.  This is the basic reason why it is worth asking how topological defect insertions affect the crossing temperature.

Let $D_a$ denote topological defect operators labelled by simple objects $a$ in a fusion category, or in a higher-categorical symmetry structure.  The fusion law has the schematic form
\begin{eqnarray}
D_aD_b &=& \sum_c N^c_{ab}D_c, \label{eq:fusion-rules}
\end{eqnarray}
where the coefficients $N^c_{ab}$ need not come from a group multiplication table and the individual $D_a$ need not be invertible.  A defect-resolved thermal trace is then
\begin{eqnarray}
Z_a(\beta) &=& \mathrm{Tr}_{\mathcal H}\left(D_a e^{-\beta H}\right), \label{eq:defect-Z} \\
F_a(T) &=& -T\log Z_a(1/T). \nonumber
\end{eqnarray}
For non-invertible defects this should not automatically be read as an ordinary positive thermal partition function.  It is safer to call $F_a$ a defect, or twisted, free energy.  In a semiclassical regime one may nevertheless organize it by saddles,
\begin{eqnarray}
Z_a(\beta) &\simeq& e^{-I_{\rm AdS,a}(\beta)}+e^{-I_{\rm BH,a}(\beta)}+\cdots . \label{eq:saddle-defect}
\end{eqnarray}
This motivates the definition of a defect-resolved Hawking--Page temperature,
\begin{eqnarray}
F_{\rm BH,a}\!\left(T^{(a)}_{\rm HP}\right) &=& F_{\rm AdS,a}\!\left(T^{(a)}_{\rm HP}\right), \label{eq:defect-HP}
\end{eqnarray}
and, when the relevant branch is well defined, a corresponding defect-resolved Davies temperature.

A small deformation makes the effect transparent.  Write
\begin{eqnarray}
\Delta F_a(T) &=& F_{\rm BH,a}(T)-F_{\rm AdS,a}(T) \label{eq:defect-free-energy-shift} \\
&=& \Delta F_0(T)+\delta\Delta F_a(T), \nonumber
\end{eqnarray}
where $a=0$ is the identity sector.  To first order, the shift of the Hawking--Page temperature is
\begin{eqnarray}
\delta T^{(a)}_{\rm HP} &=& -\frac{\delta\Delta F_a(\THP)}{\partial_T\Delta F_0(T)|_{T=\THP}}. \label{eq:defect-HP-shift}
\end{eqnarray}
The induced shifts of the normalized ratios are
\begin{eqnarray}
\delta C^{(a)}_T &=& \frac{\delta T^{(a)}_{\rm HP}-R_T\,\delta T^{(a)}_{\rm D}}{\Tmin}, \label{eq:defect-ratio-shifts} \\
\delta C^{(a)}_S &=& \frac{\delta S^{(a)}_{\rm HP}-R_S\,\delta S^{(a)}_{\rm D}}{\Smin}, \nonumber
\end{eqnarray}
with $R_T=\THP/\Tmin$ and $R_S=\SHP/\Smin$.  The ratios are therefore protected only in the special case where the defect moves the Hawking--Page and Davies data by the same relative scale.  If the defect weights the two saddles differently, the topological dipole splits into defect sectors.

Fusion gives an additional constraint.  From \Eq{eq:fusion-rules},
\begin{eqnarray}
Z_{D_aD_b}(\beta) &=& \mathrm{Tr}\left(D_aD_b e^{-\beta H}\right) \label{eq:fusion-Z} \\
&=& \sum_c N^c_{ab} Z_c(\beta). \nonumber
\end{eqnarray}
This is a linear statement for partition functions.  The corresponding free energies obey a log-sum-exp relation, and reduce to a dominant-sector relation only after a saddle approximation has been made.  This distinction is important: the fusion algebra constrains the family of defect-resolved Hawking--Page temperatures, but it does not imply a linear relation among the free energies themselves.

One concrete testing ground is a thermal orbifold or discrete gauging.  In such examples the low-temperature saddle may acquire a Casimir or twisted-sector contribution, while the black-hole saddle can respond differently because the thermal circle is contractible.  Holographic examples with non-invertible symmetries in orbifold CFTs and brane constructions provide natural backgrounds in which to check this idea \citep{ApruzziBahBonettiSchaferNameki2023,GutperleLiRathoreRoumpedakis2024,AntinucciBeniniCopettiGalatiRizi2025}.  Matrix-model descriptions of deconfinement, where the transition is controlled by the Polyakov-loop eigenvalue distribution, offer another useful language \citep{CopettiGrassiKomargodskiTizzano2022}.

A compact dictionary can be stated in words.  The Hawking--Page temperature is the zero of $\F$ with $w_{\rm HP}=+1$, and in a defect sector it is replaced by the equality of the defect-resolved saddle free energies.  The Davies temperature is the heat-capacity divergence with $w_{\rm D}=-1$, and may become a sector-dependent stability threshold.  The ratios $C_T$ and $C_S$ are normalized first moments of the pair, while in a sector $a$ they become $C_T^{(a)}$ and $C_S^{(a)}$ if the two saddles are shifted differently.  Finally, thermal AdS and the AdS black hole differ in the contractibility of the thermal circle: a defect can remain as a non-contractible twist in one saddle and be absorbed, condensed, or reorganized in the other.  This is the point where fusion data can enter.

We can identify three working hypotheses:  first, if a defect contributes the same phase-independent factor to both saddles, the ratios should remain unchanged even if individual scales move.  Second, if the defect distinguishes the saddles, the Hawking--Page temperature may split into sector-dependent values constrained by \Eq{eq:fusion-Z}.  Third, in phase diagrams with several transition points, the moment hierarchy of \SecRef{sec:multipoles} may itself carry a representation or module structure under the relevant categorical symmetry.  Establishing any of these statements is beyond the scope of this work, it would require a top-down model in which both the black-hole saddles and the defect sectors are under analytic control.

\section{Discussion, conclusions and outlook}
\label{sec:conclusion}

In summary, the universal Hawking--Page ratios $C_S$ and $C_T$ can be written in a compact topological form.  In the common vector-field construction the Davies point and the Hawking--Page point carry winding numbers $-1$ and $+1$.  The total charge of the pair is therefore zero, but the pair has a non-zero signed first moment.  In entropy and temperature variables,
\begin{eqnarray}
\Pmom_S &=& \SHP-\Smin, \\
\Pmom_T &=& \THP-\Tmin, \nonumber
\end{eqnarray}
and normalizing these by the Davies-point scales gives
\begin{eqnarray}
C_S &=& \frac{\SHP-\Smin}{\Smin}, \\
C_T &=& \frac{\THP-\Tmin}{\Tmin}. \nonumber
\end{eqnarray}
This is the sense in which the Hawking--Page/Davies pair behaves as a thermodynamic topological dipole.

For four-dimensional Schwarzschild--AdS and grand-canonical Reissner--Nordstr\"om--AdS black holes the familiar values are recovered,
\begin{eqnarray}
C_S &=& 2, \\
C_T &=& \frac{2}{\sqrt{3}}-1. \nonumber
\end{eqnarray}
For charged non-rotating AdS black holes in $d$ spacetime dimensions one obtains
\begin{eqnarray}
C_S(d) &=& \left(\frac{d-1}{d-3}\right)^{(d-2)/2}-1, \\
C_T(d) &=& \frac{d-2}{\sqrt{(d-1)(d-3)}}-1. \nonumber
\end{eqnarray}
At fixed angular velocity in four-dimensional Kerr--AdS the winding numbers remain unchanged, while the normalized separations acquire rotational corrections beginning at order $\Omega^4$.

The same geometry of the free-energy curve also identifies the Davies point as the top of the Hawking--Page barrier.  This gives the dimensionless barrier
\begin{eqnarray}
B &=& \frac{\F(\Smin)}{\Tmin\Smin},
\end{eqnarray}
with $B=1/3$ in four dimensions and $B(d)=1/[(d-1)(d-3)]$ in the higher-dimensional charged non-rotating family.  Together with the shape-function argument, this explains why pressure, electric potential, and other simple scales can drop out of the ratios: they rescale the curve without changing its reduced shape.

The dipole picture  does not replace the ordinary thermodynamic calculation of $\Smin$, $\SHP$, $\Tmin$, and $\THP$, 
the moments mimick  topological invariants in a more  strict mathematical sense.  It does, however, put the universal constants and the winding numbers into one language.  That language becomes useful once the phase diagram contains several transition points, because the same construction extends to a moment hierarchy.  Higher moments can distinguish diagrams that have the same total winding number and even the same first moment.

We have proposed a categorical approach  which could be of interest for future work.  If topological defects associated with generalized or non-invertible symmetries can be inserted into the thermal trace, then the Hawking--Page temperature can be resolved by defect sector.  Fusion rules then constrain the allowed pattern of sector-dependent temperatures and, consequently, the sector-dependent dipole ratios.  A concrete test would be to compute these quantities in a holographic orbifold, brane realization, or matrix-model description where both the thermal AdS and black-hole saddle contributions are known.  Such a calculation would decide whether the dipole ratios are protected, shifted, or split by categorical data.

\section*{Acknowledgments}

The author acknowledges financial support from the Spanish government MEC/2025-27468 and CARM/2025-76885 research projects.  AI-assisted language editing has been used  to improve the general style and to check mathematical formulas.

\end{document}